\documentstyle[prl,aps,preprint,epsfig]{revtex}
\begin{document}
\tighten
\preprint{UMD PP\#01-007 \\ DOE/ER/40762-214}
\title{STUDYING HADRONIC STRUCTURE OF THE PHOTON \\ IN LATTICE QCD}
\author{Xiangdong Ji and Chulwoo Jung}
\address{Department of Physics, 
University of Maryland, 
College Park, Maryland 20742 }
\date{Jan. 22, 2001}

\maketitle

\begin{abstract}
We show that the matrix element of a local 
quark-gluon operator in the
photon state, $\langle \gamma(k\lambda)|\hat O|
\gamma(k\lambda)\rangle$, can be calculated in lattice QCD. 
The result is generalized to other quantities 
involving space-like photons, 
including the transition form factor $\gamma\gamma^*\rightarrow \pi^0$ 
and the virtual-photon-nucleon 
Compton amplitude $\langle \gamma^*N |\gamma^*N\rangle$ 
which can be used to define the generalized 
Drell-Hearn-Gerasimov and Bjorken sum rules.
\end{abstract}
\vspace{1in}

\narrowtext
Ever since the 1960's, it has been well known that the 
photon is not just a point-like particle; rather, it 
has a complicated hadronic structure. Similarity of
the photon-nucleon scattering cross section to that of
meson-nucleon scattering has led to the so-called
vector dominance model of photon-hadron interactions \cite{sakurai}. 
In recent years, the photon structure has been explored 
extensively in $e^+e^-$ and $ep$ collider experiments 
\cite{nisius}.
For instance, in the certain kinematic region, 
$e^+e^-$ scattering is effectively the real and deeply-virtual 
photon collision, from which the partonic structure 
of the real photon can be extracted. Indeed, 
the unpolarized quark and gluon distributions in the photon 
has been phenomenologically determined to a reasonable
accuracy \cite{nisius}, and future data from the 
polarized HERA and $e^+e^-$ colliders can constrain the 
polarized distributions as well \cite{plan}. 

Given the large amount of experimental data 
available to describe the hadronic properties 
of the photon, it is appropriate to ask  
how they can be understood from the fundamental theory---
quantum chromodynamics (QCD). To the authors' knowledge,
the answer is still open. Part of the reason is that 
the photon is not an eigenstate of QCD; the 
standard lattice QCD method used in calculating the
matrix elements of the nucleon and pion does not apply 
\cite{liu}. As we shall explain, the QCD
matrix elements in the photon state are time-dependent 
correlations which are notoriously difficult to access
through the Euclidean space \cite{maiani}. However, we will show 
in this paper that they can be extracted
from the Euclidean correlation functions
on a lattice. In particular, the moments 
of polarized (unpolarized) quark and gluon distributions 
of the photon can be calculated in Monte Carlo 
simulations just like for those of the nucleon. 
Generalizing our finding, we show that a number of 
processes involving space-like virtual photons, such as 
transition $\gamma\gamma^* \rightarrow \pi$ \cite{brodsky} 
and the virtual-photon-nucleon 
forward Compton scattering $\gamma^* N
\rightarrow \gamma^* N$, can be studied in lattice QCD.
The amplitude for the latter process is the key
for generalizing the well-known Drell-Hearn-Gerasimov
and Bjorken sum rules for the spin-dependent
structure function of the nucleon \cite{jiosborne}.

We start with a simple but instructive discussion
about what is meant by the hadronic or QCD structure of the
photon. If $\hat O(0)$ is a quark-gluon operator at the 
spacetime point 0, its matrix element in the photon state can be
defined from the standard 
Lehmann-Symanzik-Zimmermann (LSZ) reduction formula, 
\begin{equation}
   \langle \gamma(k\lambda')|\hat O(0)|\gamma(k\lambda)\rangle
  = -\lim_{k'^\mu\rightarrow k^\mu} \int d^4xd^4y e^{-ik'(y-x)}
   k'^2\langle 0|T\epsilon^*(\lambda')\cdot 
    A(x) ~\hat O(0)~ \epsilon(\lambda)\cdot A(y)
  |0\rangle k'^2  \ , 
\end{equation}
where $k^\mu$ is the on-shell ($k^2=0$) photon momentum, 
$\lambda$ and $\lambda'$ are the photon helicities, 
and $k'^\mu$ on the right-hand side is taken to 
be on-shell after cancelling the photon poles. The normalization
for the photon states are taken to be covariant, i.e.,
$\langle \gamma(k'\lambda')|\gamma(k\lambda)\rangle
  =2k^0(2\pi)^3\delta(\vec{k}'-\vec{k})\delta_{\lambda\lambda'}$. 
The renormalization constant $Z_3= 1+{\cal O(\alpha_{\rm em})}$ 
has been omitted for simplicity. Evaluating the 
electromagnetic part of the Green's function in  
the lowest-order perturbation theory (this is allowed
because of the small electromagnetic coupling), we find
\begin{equation}
   \langle \gamma(k\lambda')|\hat O(0)|\gamma(k\lambda)\rangle
  =  -e^2 \int d^4xd^4y e^{-ik(y-x)}
 \epsilon^{*\mu}(\lambda') \langle 0|TJ_{\mu}(x) 
\hat O(0) J_\nu(y)|0\rangle \epsilon^\nu(\lambda) \ ,  
\label{mresult}
\end{equation}
where $e$ is 
the renormalized electric charge unit and $J^\mu$ is the
electromagnetic current operator of the quarks. According to the
above equation, the matrix element of interest
is a Minkowski-space correlation function in QCD. 
This is in contrast to an analogous nucleon matrix element, 
$\langle p|\hat O|p\rangle$, which has no explicit reference 
to the Minkowski time. 

One can remove the explicit time dependence
by deriving a Lehmann representation of Eq. (\ref{mresult}):
separating different time-orderings, inserting complete
sets of hadronic states, and integrating out the 
spacetime coordinates 
$x$ and $y$. The result can be shown as the 
time-independent perturbation diagrams in Fig. 1. 
Diagrams a) to f) represent contributions 
from six possible time-orderings. 
Diagrams g) to i) are the
vacuum contributions which must be subtracted.
As an example, the diagram a) corresponds to the
following time-independent matrix element,
\begin{equation}
\langle \gamma(k\lambda')|\hat O(0)|\gamma(k\lambda)\rangle_{\rm Fig.~1a}
= e^2 \sum_{nn'}
   {\langle 0 |\epsilon^*(\lambda')\cdot J(\vec{k})|n'\rangle
   \langle n'|\hat O(0)|n\rangle \langle n|\epsilon(\lambda)
\cdot J(\vec{k})|0\rangle
   \over (\omega-E_n)(\omega-E_{n'})}  \ , 
\label{lehmann}
\end{equation}
where $|n\rangle$ are the  
eigenstates of $H_{\rm QCD}$ with momentum $\vec{k}$, 
and $\omega =| \vec{k}|$ is the on-shell
photon energy. The contributions from the other diagrams
can be expressed similarly using the rules 
from time-independent perturbative theory. 

\begin{figure}[t]
\begin{center}
\epsfig{file=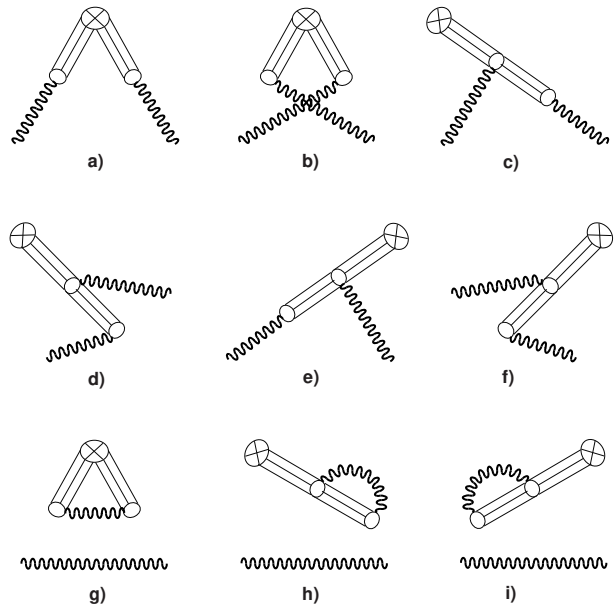,clip=,height=8cm,angle=0}
\vspace{0.3in}
\end{center}
\caption{Time-ordered diagrams for the quark-gluon matrix elements
of the photon state. The circles with crosses represent
local quark and gluon operators, and the open ones
the electromagnetic currents. Multiple parallel
lines indicate the intermediate hadronic eigenstates.}
\label{fig1}
\end{figure}   

Though hardly new, the Lehmann representation helps us in 
understanding the physical origin of the vector dominance. 
With certain probability, the photon is a superposition 
of a set of hadron states with the photon quantum numbers. 
The lowest-lying such states are $\pi\pi$ scattering states
in the $p$-wave. A strong resonance in this channel is 
the $\rho$ meson. The vector dominance then refers to  
the dominance of the $\rho$ resonance in 
the above hadron-state summations, although
other hadron states can and do contribute. 
It is interesting to note from Fig. 1
that the hadronic part of the photon wave function 
does not consist of just the pure hadron states; 
it contains also the components that are the mixtures
of one or more resolved photons and hadrons. 

The above discussion makes it clear that
the hadronic structure of the photon 
is qualitatively different from that of ordinary
hadrons, at least in the context of QCD. 
The photon matrix elements contain explicit
sums over the complete set of QCD eigenstates 
as well as the contributions from different time-orderings. 
Because the time-dependent correlation functions
are generically difficult to study in 
the Euclidean space, there are no general
rules to calculate them in lattice QCD. 
One might be tempted to use the electromagnetic 
current $J^\mu$ as a photon interpolating field 
to extract the matrix element of interest from  
$\langle 0|J_\mu(y) \hat O(0) J_\nu(x)|0\rangle$ at 
large positive $-x_4, y_4$. The actual outcome is a 
matrix element between the threshold 
two-$\pi$ states, or the $\rho$ if it were stable.

The key observation for a lattice calculation of 
the photon matrix element is that the photon is 
an eigenstate of the combined eletromagnetic and 
strong interaction hamiltonian \cite{muller}. When both 
interactions are present, we can use the electromagnetic 
potential $A^\mu$ as the interpolating field for photon. 
Thus we consider the following matrix elements in the 
Euclidean space,
\begin{equation}
    \langle 0|A_\mu(y)\hat O(0) A_\nu(x)|0\rangle \ , 
\end{equation}
where $x$ and $y$ are Euclidean coordinates. After 
Fourier-transforming the spatial coordinates 
and letting $y_4$ and $-x_4$ be large, we have
\begin{eqnarray}
  M_{\mu\nu}(&y_4&,x_4,\vec{k})  =  
   \int d^3\vec{x} e^{-i{\vec k} \cdot\vec{x}}
   \int d^3\vec{y} e^{i\vec{k}\cdot \vec{x}}
\langle 0|A_\mu(\vec{y}, y_4) \hat O(0) A_\nu(\vec{x}, x_4)|0\rangle
   \nonumber \\
&& \rightarrow \sum_{\lambda\lambda'} \langle 0|A_\mu(\vec{k})|\gamma(k\lambda')\rangle
   e^{-y_4 \omega} \langle \gamma(k\lambda')|\hat O(0)|
   \gamma(k\lambda)\rangle e^{x_4 \omega} \langle \gamma(k\lambda)| 
    A_\nu(\vec{k})|0\rangle \ . 
\label{above}
\end{eqnarray}
The summations over $\lambda$ and $\lambda'$ can be selected
by multiplying $M_{\mu\nu}$ with appropriate polarization
vectors. In the second line, we have neglected 
the three-photon states with the same momentum which
are suppressed by powers of $\alpha_{\rm em}$, and 
the hadronic states which are suppressed by an energy 
gap ($2m_\pi$ at $\vec{k}\sim 0$). 

Before putting $M_{\mu\nu}(y_4,x_4, \vec{k})$ on a lattice, 
we integrate out the photon field 
in perturbation theory, 
\begin{eqnarray}
   M_{\mu\nu}(y_4, x_4, \vec{k})
  &=& e^2\int d^4wd^4z e^{-i\vec{k}\cdot(\vec{w}-\vec{z})} 
   D_{\mu\rho}(\vec{k},y_4-w_4) \nonumber \\ && \times
\langle 0 |T_E J_\rho(\vec{w}, w_4) \hat O(0) 
J_\sigma(\vec{z}, z_4)|0
\rangle D_{\sigma\nu}(\vec{k},z_4-x_4) \ ,
\end{eqnarray}
where $D^{\mu\nu}$ is the photon propagator 
in the Euclidean space and $T_E$ is the Euclidean 
time-ordering. Using
the Fourier integral ($k>0$), 
\begin{equation}
   \int^\infty_{-\infty} 
   d\omega {e^{i\omega z}\over \omega^2 + k^2}
     = {\pi\over k} e^{-k|z|} \ , 
\end{equation} 
we can write 
\begin{equation}
   M_{\mu\nu}(y_4,x_4,\vec{k})
 \sim \int dw_4dz_4 e^{-\omega|z_4-x_4|} 
    e^{-\omega|y_4-w_4|} \langle 0 |T_E J_\mu(-\vec{k},w_4)
   \hat O(0) J_\nu(\vec{k},z_4)|0\rangle \ . 
\end{equation}
Taking the limit $y_4, -x_4 \rightarrow \infty$
and comparing the result with Eq. (\ref{above}),  
we get, 
\begin{equation}
    \langle \gamma(k\lambda')|\hat O(0)|
    \gamma(k\lambda)\rangle = e^2\int dw_4 dz_4
         e^{\omega (w_4-z_4)}
       \epsilon^*_i(\lambda')\langle 0 |T_E J_i(-\vec{k},w_4)
   \hat O(0) J_j(\vec{k},z_4)|0\rangle
        \epsilon_j(\lambda) \ ,
\label{eresult}
\end{equation}
where we have restored the photon polarization in physical gauge
($\epsilon^0=0$ and $\vec{\epsilon}\cdot\vec{k}=0$), $i$ and $j$
sum over spatial indices, and $J_\mu = \sum_f e_f \bar 
\psi_f\gamma_\mu\psi_f$ is the electromagnetic current summing 
over all quark flavors $f$. The above equation is one of the 
main results of this paper. 

Several comments on Eq. (\ref{eresult}) are as follows: 
First, the Euclidean correlation function
$\langle 0|T_EJ^\mu(\vec{k}, w_4) \hat O(0) 
J^\nu(\vec{k}, z_4)|0\rangle$ can be calculated
in the standard Monte Carlo simulations.  
Hence, the photon matrix elements are
accessible in lattice QCD just like the nucleon ones. 
Second, the above
expression {\it is} nothing but a straightforward 
analytical continuation of the expression
in Eq. (\ref{mresult}). However,
the process arriving at Eq. (\ref{eresult}) 
makes it clear that the analytical continuation
fails if there is no energy gap between the
photon and hadronic states. Finally, $\langle 0|TJ_\mu(\vec{k}, 
w_4) \hat O(0) J_\nu(\vec{k}, z_4)|0\rangle$ 
is what one would naively consider when 
using $J^\mu$ as a photon interpolating field. 
On the other hand, the correct result involves integrations
over $w_4$ and $z_4$, which include not only
the lowest-energy hadron state but all others
with the photon quantum numbers.
Finally, it is straightforward to show that the
above Euclidean correlation function reproduces
the Lehmann representation of the photon matrix 
element. In the process, one finds
that if the energy of the photon exceeds that of the
lowest hadronic state (the gap vanishes), 
Eq. (\ref{eresult}) is divergent. 
Thus, we emphasize that the photon
matrix element is calculable on a lattice 
only when there is an  
energy gap between the photon and hadron states.

In the remainder of the paper, we explore under
what conditions the above result can be 
generalized. First of all, the initial
and final photons can have different 
3-momenta, and then the forward matrix element 
is generalized to form factors. Next, 
we introduce the hadronic structure 
of an off-shell photon for which $\omega \ne |\vec{k}|$. 
The parton distributions in an off-shell photon 
can be extracted from experimental data just as
for an on-shell photon \cite{nisius}. In QCD, the 
off-shell photon matrix elements 
are defined in terms of the same Minkowski correlation 
in Eq. (\ref{mresult}). The corresponding 
Euclidean expression in Eq. (\ref{eresult})
is valid only when $0<\omega<\sqrt{4m_\pi^2 + \vec{k}^2}$, 
where the upper limit refers to 
the hadron production threshold for a time-like
photon. Finally, one can define and calculate on 
a lattice the hadronic matrix elements of the 
weak-interaction gauge bosons, off- or on-shell, 
by replacing in Eq. (\ref{eresult}) 
the electromagnetic currents with the weak-interaction currents.

To make further progress, we explore under 
what conditions a Minkowski correlation can
be obtained from a numerical calculation of 
the corresponding Euclidean correlation. Euclidean 
Green's functions are generally real as they can be 
expressed as real integrals in the functional 
formulation and amenable to Monte Carlo simulations,
whereas Minkowski correlations can 
have imaginary parts corresponding to physical 
intermediate states propagating infinite 
distances. Therefore, it is reasonable to expect
that Minkowski correlations can be analytically continued 
and calculated in Euclidean space only when they are 
in kinematic domains where no physical intermediate states are
allowed. In the example of the matrix elements
for an off-shell photon, the condition $\omega<
\sqrt{4m_\pi^2+\vec{k}^2}$ ensures that the photon
can never turn into a real hadron state.

Thus the transition form factor (amplitude) corresponding to 
$\gamma\gamma^*\rightarrow \pi^0$ is calculable on a lattice, 
where the first photon is real and the second is 
space-like. In Minkowski
space the amplitude can be expressed as 
\begin{equation}
    A_{\mu\nu}(P, k) = \int d^4y 
     e^{iky}\langle 0|T J_\mu(y)J_\nu(0)|\pi(P)\rangle \ . 
\end{equation}
Denoting the 
interpolating field for the pion as $\phi$, we construct
the following Green's function in the Euclidean space, 
\begin{equation}
   G_{\mu\nu}(\vec{P},x_4,k)
  = \int dy_4 e^{y_4\omega}
   \langle 0|T_E J_\mu(\vec{k},y_4) J_\nu(0)
             \phi(\vec{P}, x_4)|0\rangle \ . 
\end{equation}
Taking $x_4\rightarrow -\infty$ and keeping only the 
pion state of momentum $\vec{P}$,  
we have
\begin{equation}
 G_{\mu\nu}(\vec{P},x_4,k)
 = \int dy_4 e^{y_4\omega}
    \langle 0|T_E J_\mu(\vec{k},y_4) J_\nu(0)|\pi(P)\rangle
         e^{x_4E_P} \langle \pi(P)|\phi(\vec{P})|0\rangle \ , 
\end{equation}
where $E_P=\sqrt{\vec{P}^2+m_\pi^2}$ is the on-shell
pion energy. By working out its Lehmann representation, 
the Euclidean correlation function reproduces 
the corresponding Minkowski amplitude. 

The final example involving the photons is the
forward virtual-photon-nucleon Compton scattering amplitude,
\begin{equation}
  T_{\mu\nu}(P,q) = i\int e^{iqx} d^4x 
       \langle P|TJ_\mu(x)J_\nu(0)|P\rangle \ ,
\end{equation}
where $|P\rangle^\mu$ is the nucleon state with momentum
$P$. The space-like virtual photon momentum
has two distinct kinematic regions, $(P+q)^2>M_N^2$
and $(P+q)^2<M_N^2$, where $M_N$ is the nucleon mass.
In the former region, the amplitude is complex 
and its imaginary part corresponds to the inclusive
production cross section in $eP$ scattering. In the
other region, the Compton amplitude is real
and therefore we expect it is calculable 
in lattice QCD. Define the four-point Euclidean
correlation, 
\begin{equation}
  G_{\mu\nu}(\vec{P},y_4,z_4,\vec{q})
  = \int dx_4e^{x_4\omega}\langle 0|T_E 
   \eta(\vec{P},y_4) J_\mu(\vec{q},x_4)J_\nu(0)\bar\eta(\vec{P},z_4)
      |0\rangle \ , 
\end{equation}
where $\eta$ is an interpolating field for the nucleon.
Take the limit $y_4,-z_4 \rightarrow \infty$, we have the
nucleon state dominance,
\begin{equation}
     G_{\mu\nu}(\vec{P},y_4,z_4,\vec{q})
  = \int dx_4e^{x_4\omega}\langle 0|\eta(\vec{P})|P\rangle
     e^{-y_4E_P}\langle P|T_E 
   J_\mu(\vec{q},x_4)J_\nu(0)|
     P\rangle e^{z_4E_P} \langle P|\bar\eta(\vec{P})
      |0\rangle \ . 
\end{equation}
Integrating over $x_4$, we find
that the remaining correlation function reproduces
the physical Compton amplitude in the region 
$(P+q)^2<M_N^2$. 

A particularly interesting application is when $\omega=0$. The
spin-dependent Compton amplitude as a function of $Q^2=-q^2$ defines
an integral over $\nu$ for the $G_1(\nu, Q^2)$ 
structure function of the nucleon. In the limit
$Q^2\rightarrow 0$, the amplitude can be obtained
by the low-energy theorem and chiral perturbation
theory and the integral defines the 
Drell-Hearn-Gerasimov sum rule and its generalization \cite{jiosborne}.
In the limit $Q^2\rightarrow\infty$, the amplitude
can be studied in perturbative QCD and the integral
defines the Bjorken sum rule and its generalization.
Our result here allows a lattice QCD
calculation of the amplitude at intermediate $Q^2$ 
where there is no firm theoretical prediction available. 

In conclusion, we have shown that the QCD
structure of the photon can be studied on a lattice
by calculating Euclidean correlation functions
involving local quark-gluon operators and electromagnetic
currents. Also calculable are a number of 
other interesting examples involving photons, 
such as $\gamma\gamma^*\rightarrow \pi^0$ and 
$\gamma^*N\rightarrow \gamma^*N$. These results 
shall lead to more exciting interplay between lattice
QCD and strong-interaction phenomenology.

\acknowledgements
X. Ji thanks A. Mueller for an interesting discussion
which led the subject of this paper and for his conviction
that the partonic structure of the photon must
be calculable in lattice QCD.
This work is supported in part by funds provided by the
U.S. Department of Energy (D.O.E.) under cooperative agreement
DOE-FG02-93ER-40762.

\end{document}